\documentclass[aps,pra,superscriptaddress,amsmath,amssymb,preprintnumbers,showpacs]{revtex4}

\usepackage{amsmath}
\usepackage{amstext}
\usepackage{latexsym}
\usepackage{graphicx}
\usepackage{amsfonts}

\begin{document}

\newcommand{\bra}[1]{\langle #1 \vert}
\newcommand{\ket}[1]{\vert #1 \rangle}
\newcommand{\scl}[2]{( #1 \vert #2 )}
\newcommand{\dert}[1]{\frac{d}{dt}#1}
\newcommand{\dertn}[2]{\frac{d^{#2}}{dt^{#2}}#1}
\newcommand{\derpt}[1]{\frac{\partial}{\partial t}#1}
\newcommand{\derpn}[3]{\frac{\partial^{#2}}{\partial {#3}^{#2}}#1}
\newcommand{\med}[1]{\langle#1\rangle}
\newcommand{\mte}[1]{\| #1 \|}
\newcommand{\abs}[1]{\vert #1 \vert}
\newcommand{\sa}[1]{{#1}_{\Sigma}}
\newcommand{\tr}[2]{tr_{#2}\big{\{}#1\big{\}}}
\newcommand{\Tr}[1]{tr \big{\{}#1\big{\}} }
\newcommand{\defi}[1]{\stackrel{def.}{=} #1 }

\title{Density matrix operatorial solution of the non--Markovian Master
Equation for Quantum Brownian Motion}

\author{F. Intravaia}
\author{S. Maniscalco}
\email{sabrina@fisica.unipa.it}
\author{A. Messina}

\affiliation{MIUR, INFM and Dipartimento di Scienze Fisiche ed
Astronomiche, via Archirafi 36, 90123 Palermo,Italy}

\date{\today}

\begin{abstract}
An original method to exactly solve the non-Markovian Master
Equation describing the interaction of a single harmonic
oscillator with a quantum environment in the weak coupling limit
is reported. By using a superoperatorial approach we succeed in
deriving the operatorial solution for the density matrix of the
system. Our method is independent of the physical properties of
the environment. We show the usefulness of our solution deriving
explicit expressions for the dissipative time evolution of some
observables of physical interest for the system, such as, for
example, its mean energy.
\end{abstract}


\maketitle

\section{Introduction}
The problem of the quantum dynamics of a small system interacting
with its environment has been extensively studied since the origin
of quantum mechanics. However, in spite of the noticeable
progresses in the theory of open quantum systems,  many conceptual
difficulties  still remain. Indeed the understanding of the
effects of the environment on the physical system of interest, in
general, is not an easy task. The conventional way to afford the
problem of the description of the time evolution of an open
quantum system consists in eliminating the degrees of freedom of
the environment in order to derive an equation of motion for the
reduced density matrix of the small  system. This procedure stems
from the fact that we are usually only interested in the dynamics
of this subsystem and do not really care about the state of the
environment and its evolution. The procedure of tracing over the
environmental variables thus amounts at declaring that everything
other than the small system is indifferent to us. In this way one
reduces the number of variables one needs to take into account to
study the problem.

Only very few physical systems prove to be amenable to an analytic
description of their open dynamics. In this paper we deal with the
simplest open quantum system worth being studied: the damped
harmonic oscillator. Such a system is of great conceptual
importance because it provides a simple, successful starting point
in the theoretical description of many experimental situations  in
the quantum optics, solid state physics and quantum field theory
contexts. For this reason it is also one of the most extensively
studied physical systems
\cite{fkm65,fk87,legget,fvc,lindenberg,haake,barnett}.

Understanding the dissipative behaviour of open quantum systems is
a crucial problem for both fundamental and applicative reasons. On
one hand, indeed, during the last few years, the interpretation of
Zurek, associating the quantum to classical transition with
environment induced decoherence effects, has become increasingly
popular. From this point of view, studying the open system
dynamics of exemplary quantum systems, i.e. harmonic oscillator,
would help in identifying the elusive border between quantum and
classical descriptions of the world. On the other hand, the huge
 advances in experimental techniques for controlling the evolution
of quantum systems have paved the way to the realization of the
first quantum logic gates, the key elements of quantum computers.
The biggest obstacle to building quantum computers is the
 decoherence process due to the unavoidable coupling with the
 external environment. Thus, studying a decohering quantum
 computer is an essential step for the identification of realistic
 quantum error correcting codes \cite{horo}.

A standard method to describe the effects of the environment on a
small quantum system is based on the Born-Markov Master Equation.
Such an approach is valid whenever the environmental correlation
time is much shorter than the typical time scales of the system
dynamics. Under these conditions, it is possible to derive and, in
some cases, to solve analytically the Master Equation ruling the
dynamics. This is for example the case of a quantum harmonic
oscillator coupled through bilinear interactions to a reservoir of
harmonic oscillators. This model is a particular case of the
Quantum Brownian Motion (QBM), where we specify the system to be
just a quantum harmonic oscillator \cite{leshouches}.

There exist situations, however, wherein the reservoir correlation
time is longer than the system time scales of interest and thus
the Born--Markov approximation does not hold anymore. This is for
instance the case of atoms decaying in photonic band gap materials
or atom lasers \cite{fotolaser}. Moreover, very recently, the
potential interest of non--Markovian reservoirs for quantum
information processing has been demonstrated \cite{Ahn} and a
non--Markovian description of quantum computing, showing the
limits of the Markovian approach, has been presented \cite{horo}.
Several procedures have been developed to treat non--Markovian
processes. Most of them are reviewed in \cite{Dalton}. Very often,
in order to describe the dynamics of an open quantum system in
terms of a master equation for its density matrix, a number of
approximations involving the relevant system and reservoir
parameters are required. One of the most common is the weak
coupling approximation based on the assumption of a weak
system-environment coupling strength. As underlined by Paz and
Zurek in \cite{leshouches}, a perturbative approach in the
coupling strength can always be shown to lead to Master Equations
local in time. This fact could appear surprising at a first sight
since often non--Markovian quantum systems are described by
generalized Master Equations involving a non--local memory kernel
taking into account the past history of the reduced system. The
advantage of dealing with a generalized Master Equation which is
local in time is that the memory effects of the environment are
incorporated in its time dependent coefficients
\cite{tcl,petruccionebook,petruccione}. This circumstance makes it
possible, in some cases, to find an analytic solution of the
generalized Master Equation.

In this paper we present an original method for the derivation of
the operatorial density matrix solution of the non--Markovian
Master Equation for QBM in the weak coupling limit. We apply our
procedure to the case in which the system is a harmonic oscillator
interacting with the quantized reservoir through a bilinear
coupling. Our method does not rely on any assumption other than
the weak coupling limit. In particular it is independent of the
type of environment considered and on the characteristic
parameters of the system, provided a perturbative approach to the
second order in the coupling constant is valid. We demonstrate the
usefulness and simplicity of our analytic solution discussing, as
an example, the dissipative dynamics of the mean energy of the
harmonic oscillator.

The paper is organized as follows. In Section II we introduce the
mathematical formalism of superoperators and we present some key
properties for solving the generalized Master Equation for our
system. In Section III we describe the non--Markovian Master
Equation for QBM in the weak coupling limit and in Section IV we
derive its operatorial solution. In Section V we discuss some
approximated forms of the density matrix and we present some
applications . Finally in section VI we present conclusions.

\section{Superoperator formalism}

Let us begin introducing the mathematical formalism and properties
which we  use in the rest of the paper to derive the time
dependence of density matrix of the system under scrutiny. Our
method is based on the extension of the notion of eigensolution of
an operator in the
 \lq\lq \textit{superoperator}\rq\rq formalism.\\

Given two generic operators $\hat{A}$ and $\hat{\rho}$, let us
define the \textit{superoperators of the S-type $\mathbf{A}^S
\equiv (A)^S$ and of the $\Sigma$-type $\mathbf{A}^{\Sigma}\equiv
(A)^{\Sigma}$} in the following way:
\begin{subequations}
\label{super}
\begin{equation}
\label{superc} \mathbf{A}^S\mathbf{\rho}=[\hat{A},\hat{\rho}],
\qquad \text{$S$ type}
\end{equation}
\begin{equation}
\label{supera}
\mathbf{A}^{\Sigma}\hat{\rho}=\{\hat{A},\hat{\rho}\}, \qquad
\text{$\Sigma$ type}
\end{equation}
\end{subequations}
where the squared  and  curl brackets indicate the commutator and
anti-commutator respectively. Eqs. \eqref{superc} and
\eqref{supera} define a particular class of superoperators
hereafter also called commutator ($S$) and anticommutator
($\Sigma$) superoperators. It is important to note that a
$S$-superoperator $\mathbf{A}^S$ may be linearly combined with or
multiplied by a $\Sigma$-superoperator $\mathbf{B}^{\Sigma}$
giving rise to superoperators belonging neither to the $S$ class
nor to the $\Sigma$ class. From Eqs. \eqref{superc} and
\eqref{supera} linearity follows immediately
\begin{equation}
\left(\alpha\hat{A}+\beta\hat{B}\right)^{S(\Sigma)}=\alpha\mathbf{A}^{S(\Sigma)}+\beta\mathbf{B}^{S(\Sigma)}
\qquad \alpha,\beta \in \mathbb{C},
\end{equation}
and the proprieties
\begin{subequations}
\label{comrel}
\begin{equation}
[\mathbf{A}^S,\mathbf{B}^S]=[\hat{A},\hat{B}]^S,
\end{equation}
\begin{equation}
[\mathbf{A}^{\Sigma},\mathbf{B}^{\Sigma}]=[\hat{A},\hat{B}]^{\Sigma},
\end{equation}
\begin{equation}
\label{3c}
[\mathbf{A}^S,\mathbf{B}^{\Sigma}]=[\mathbf{A}^{\Sigma},\mathbf{B}^S]=[\hat{A},\hat{B}]^{\Sigma}.
\end{equation}
\end{subequations}

The main consequence of such  relations is that if the commutator
of two operators is a c-number then the corresponding $S$(
$\Sigma$)  superoperators do commute.For example, for the
super-operators $\mathbf{X}^{S(\Sigma)}$ and
$\mathbf{P}^{S(\Sigma)}$, corresponding respectively to  position
$\hat{X}$ and conjugate momentum $\hat{P}$ operators,  the
following relation holds
\begin{equation}
\label{scalar} [\mathbf{X}^{S(\Sigma)},\mathbf{P}^{S(\Sigma)}]=0.
\end{equation}
Moreover, from Eq. \eqref{3c} one gets
\begin{equation}
[\mathbf{X}^S,\mathbf{P}^{\Sigma}]=[\mathbf{X}^{\Sigma},\mathbf{P}^S]=2\imath
. \label{scalar2}
\end{equation}
Another superoperatorial identity  extensively used in the rest of
the paper is:
%\begin{subequations}
\begin{equation}
(\hat{A}\hat{B}+\hat{B}\hat{A})^S=(\mathbf{A}^S\mathbf{B}^{\Sigma}+\mathbf{B}^S\mathbf{A}^{\Sigma}),
\end{equation}
and, in particular,
\begin{equation}
\mathbf{A}^S\mathbf{A}^{\Sigma}=(\hat{A}^2)^S=\mathbf{A^2}^S.
\end{equation}
%\end{subequations}

In a certain sense superoperators can be considered as a
generalization of the concept of operator used in quantum
mechanics. Superoperators act on a space whose elements are
operators to give other operators. Let us  consider a
generalization of the concept of eigenstates. Looking at the
equations
\begin{subequations}
\label{auto}
\begin{gather}
\mathbf{X}^S\:e^{-\imath(p \hat{X}-x\hat{P})}=-x\:e^{-\imath(p \hat{X}-x\hat{P})}, \\
\mathbf{P}^S\:e^{-\imath(p
\hat{X}-x\hat{P})}=-p\:e^{-\imath(p\hat{X}-x\hat{P})}, \qquad x,p
\in \mathbb{R},
\end{gather}
\end{subequations}
we can  identify $e^{\imath(p \hat{X}-x\hat{P})}$ as
the eigenoperator of the superoperators $\mathbf{X}^S$ and $\mathbf{P}^S$ with
eigenvalues $-x$ and $-p$ respectively.\\
It is not difficult to show that the operators
$\left\{\frac{1}{\sqrt{2 \pi}} e^{\imath(p
\hat{X}-x\hat{P})}\right\}$ define a complete basis
\cite{weyl,mandelwolf} in the operatorial space  associated to the
single particle Hilbert space. Such a basis is orthonormal with
respect to the scalar product between two generic operators $A$
and $B$, defined as follow
\begin{equation}
\scl{\hat{A}}{\hat{B}}\defi{\Tr{\hat{A}^{\dag}\hat{B}}},
\end{equation}
since
\begin{equation}
\label{ort} \frac{1}{2\pi}\Tr{e^{\imath(p'
\hat{X}-x'\hat{P})}\:e^{-\imath(p
\hat{X}-x\hat{P})}}=\delta(x-x')\delta(p-p').
\end{equation}
It is then possible to express the density-matrix of a single
particle  in the following form \cite{weyl,mandelwolf,
glauber1,glauber2}
\begin{equation}
\label{ofe} \hat{\rho}=\frac{1}{2\pi}\int \chi(x,p)\:e^{-\imath(p
\hat{X}-x\hat{P})}\:dxdp
\end{equation}
where the scalar function $\chi(x,p)=\Tr{e^{\imath(p
\hat{X}-x\hat{P})}\hat{\rho}}$, known in the literature as
\textit{Quantum Characteristic Function} (QFC)
\cite{mandelwolf,glauber1,glauber2,gardiner}, satisfies the
following properties:
\begin{equation}
\chi(0,0)=1,\qquad\chi(x,p)=\chi^*(-x,-p),\qquad\abs{\chi(x,p)}\le
M \qquad (M \in \mathbb{R}).
\end{equation}

%The generality of those relations need some comments. The
%representation of the density matrix and all the properties which
%characterize it (at least with some simpler generalization) are
%valid for all systems completely thoroughly defined through a
%position and momentum operator or pseudo ones. Before we tacitly
%defined it referring to our aim to describe an harmonic oscillator
%but, of course, with a three dimensional generalization (and, if
%necessary, including fermionic degrees of freedom) it would be
%able to describe the state of an hydrogen atom. Here after,
%however, to our purposes, we will refer ourselves to operator an
%state Hilbert space of one dimensional harmonic oscillator.

Another superoperatorial eigenvalue equation of interest in this
paper is
\begin{equation}
\label{n} \mathbf{N}
\left(p\hat{X}-x\hat{P}\right)^n=n\left(p\hat{X}-x\hat{P}\right)^n
\qquad n \in \mathbb{N},
\end{equation}
where
$\mathbf{N}=-\frac{\imath}{2}\left(\mathbf{P}^{\Sigma}\mathbf{X}^{S}-\mathbf{X}^{\Sigma}\mathbf{P}^{S}\right)$.
In particular we have
\begin{equation}
\mathbf{N}\hat{X}^n=n\hat{X}^n, \qquad \mathbf{N}
\hat{P}^n=n\hat{P}^n.
\end{equation}
In the following we will use, for the sake of simplicity, a matrix
representation of the previous relations. Let us define
\begin{equation}
\label{vnotation} \hat{\vec{Z}}=
\begin{pmatrix}
\hat{X}\\
\hat{P}
\end{pmatrix}
\qquad \vec{\mathbf{Z}}^{S(\Sigma)}=
\begin{pmatrix}
\mathbf{X}^{S(\Sigma)}\\
\mathbf{P}^{S(\Sigma)}
\end{pmatrix}
\qquad \vec{z}=
\begin{pmatrix}
x\\
p
\end{pmatrix}
\end{equation}
With this notation Eqs. \eqref{auto}, \eqref{ort}  and \eqref{ofe}
can be recast in a more compact form
\begin{equation}
\vec{\mathbf{Z}}^S\:e^{-\imath\:\vec{z}^t\mathsf{J}\hat{\vec{Z}}}=-\vec{z}\:e^{-\imath\:\vec{z}^t\mathsf{J}\hat{\vec{Z}}},
\end{equation}
\begin{equation}
\Tr{e^{\imath\:\vec{z'}^t\mathsf{J}\hat{\vec{Z}}}
e^{-\imath\:\vec{z}^t\mathsf{J}\hat{\vec{Z}}}}=\delta[\vec{z}-\vec{z'}],
\end{equation}
\begin{equation}
\hat{\rho}=\frac{1}{2\pi}\int
\chi(\vec{z})\:e^{-\imath\:\vec{z}^t\mathsf{J}\hat{\vec{Z}}}\:d^2\vec{z},
\end{equation}
In these equations  $\mathsf{J}$ is the following $2\times 2$
matrix
\begin{equation}
\label{matrixj} \mathsf{J}=
\begin{pmatrix}
0&-1\\
1&0
\end{pmatrix},
\end{equation}
and thus satisfies the properties
$\mathsf{J}^t=\mathsf{J}^{-1}=-\mathsf{J} $ and $\mathsf{J}^2=-1$.
Note that, using this  notation, the superoperator $\mathbf{N}$
can be cast in the form
$\mathbf{N}=-\frac{1}{2}\imath\:(\vec{\mathbf{Z}}^{\Sigma})^t\mathsf{J}\vec{\mathbf{Z}}^{S}$.

\section{The Master Equation in the weak coupling limit}
In this section we specify the physical system we wish to study,
namely a single harmonic oscillator interacting with a quantized
environment, and we introduce and discuss the generalized master
equation governing its dynamics. Let us consider a harmonic
oscillator of frequency $\omega_0$ surrounded by a generic
environment. We express the total Hamiltonian $\hat{H}$ as follows
\begin{equation}
\hat{H}=\hat{H}_0 + \hat{H}_E + \alpha \hat{X}\hat{E},
\label{eq:secI1}
\end{equation}
where $\hat{H}_0=\frac{1}{2}\left(\hat{P}^2+\hat{X}^2\right)$,
$\hat{H}_E$ and $\alpha \hat{X}\hat{E}$  are the system,
environment and interaction Hamiltonians respectively and $\alpha$
is the  coupling constant. The interaction Hamiltonian here
considered has a simple bilinear form with $\hat{X}$ position
operator of the system and $\hat{E}$ generic environmental operator.\\
For the sake of simplicity we have written the previous
expressions in terms of adimensional position and momentum
operators. Let us denote with $\hat{\tilde{\rho}} \equiv
\tilde{\rho}$ the density matrix operator for the
oscillator-environment system.

%and for this reason the constant $\alpha$ contains some quantities
%related to the harmonic oscillator like the frequency $\omega_0$
%and the mass $m$: $\alpha=\sqrt{\frac{\hbar}{m \omega_0}}g$ where
%$g$ is the real coupling constant.

Let us now assume that
\begin{enumerate}
\item
at $t=0$ system and environment are uncorrelated, that is
$\tilde{\rho}(0)= \hat{\rho} (0) \bigotimes \hat{\rho}_E(0)$, with
$\hat{\rho}$ and $ \hat{\rho}_E$ density matrices of the system
and the environment respectively;
\item
$[\hat{H}_E,\hat{\rho}_E(0)]=0$ (stationarity of the environment
);
\item
$\tr{\hat{E} \hat{\rho}_E(0)}{E}=0$ (as for example in the case of
a thermal reservoir).
\item a second order perturbative approach in the coupling
constant is possible
\end{enumerate}
Under these conditions one can show \cite{rwa,leshouches} that the
non-Markovian generalized Master Equation describing the harmonic
oscillator dynamics, in the Schr\"{o}dinger picture, is the
following
\begin{equation}
\frac{d \hat{\rho}(t)}{dt} =  \left[\frac{1}{i \hbar}
\mathbf{H}_{0}^S - \mathbf{D}_S(t) + \imath
\mathbf{G}_S(t)\right]\hat{\rho}(t), \label{eqs1}
\end{equation}

where the superoperator $\mathbf{D}_S(t)$ (not $S$ or $\Sigma$
type) is defined as
%\begin{subequations}
%\label{diffdiss}
\begin{equation}
\mathbf{D}_S(t)=\int_0^t \kappa (\tau) \mathbf{X}^S \left( \cos
\omega_0\tau \mathbf{X}^S - \sin \omega_0 \tau \mathbf{P}^S
\right)d\tau, \label{eq:secIII6a}
\end{equation}
Also  $\mathbf{G}_S(t)$ is neither an $S$-type nor a $\Sigma$-type
superoperator being defined as follows
\begin{equation}
 \mathbf{G}_S(t)
= \int_0^t \mu (\tau) \mathbf{X}^{S} \left( \cos \omega_0\tau
\mathbf{X}^{\Sigma} - \sin \omega_0\tau \mathbf{P}^{\Sigma}
\right)d\tau. \label{eq:secIII6b}
\end{equation}
%\end{subequations}

In Eqs.\eqref{eq:secIII6a} and \eqref{eq:secIII6b} we have
introduced the correlation $\kappa(\tau)$ and the susceptibility
$\mu(\tau)$ functions \cite{afic}. Such quantities, characterizing
the environmental temporal behavior, are defined as follows
\begin{equation}
\kappa(\tau)= \frac{\alpha^2}{2 \hbar^2} \langle \{
\hat{E}(\tau),\hat{E}(0)\} \rangle, \label{eq:secI16a}
\end{equation}
\begin{equation}
\mu(\tau)= \frac{\imath\alpha^2}{2 \hbar^2} \langle [
\hat{E}(\tau),\hat{E}(0)] \rangle, \label{eq:secI16b}
\end{equation}

with $\langle \cdot\cdot\cdot \rangle=\tr{\cdots
\hat{\rho}_E(0)}{E}$.

The form of Eq. \eqref{eqs1} has a clear physical meaning. The
superoperator $\mathbf{D}_S(t)$, indeed, is strictly related to
diffusion (decoherence) processes only \cite{leshouches}. The
superoperator $\mathbf{G}_S(t)$, on the other hand, describes
dissipation and frequency renormalization processes
\cite{leshouches}. Such a superoperator arises from a quantum
mechanical treatment of the environment and, indeed, vanishes when
the environment is treated as a classical quantity (see also
Eq.\eqref{eq:secI16b}).

First of all, we concentrate our attention on $\mathbf{G}_S(t)$.
By using the properties of the superoperators introduced in the
previous section, one can show that such a superoperator can be
recast in the following form
\begin{equation}
\mathbf{G}_S(t) = \frac{1}{2}\left[ r(t)\hat{X}^2 -\gamma(t)\left(
\hat{X}\hat{P}+\hat{P}\hat{X}\right) \right]^S
-\imath\gamma(t)\left(\mathbf{N}+2\right), \label{dissi}
\end{equation}
where the superoperator $\mathbf{N}$ is defined by Eq.\eqref{n}
and
\begin{equation}
\label{rtgt} r(t)=2\int_0^t \mu(\tau)\cos \omega_0 \tau \:d\tau,
\qquad \gamma(t)=\int_0^t \mu(\tau)\sin \omega_0 \tau \:d\tau.
\end{equation}
As for $\mathbf{D}_S(t)$, it is straightforward to see from Eq.
\eqref{eq:secIII6a} that it can be recast in the form
\begin{equation}
\mathbf{D}_S(t)=\bar{\Delta}(t)(\mathbf{X}^S)^2-\Pi(t)\mathbf{X}^S\mathbf{P}^S,
\end{equation}
where
\begin{equation}
\bar{\Delta}(t)=\int_0^t\kappa(\tau)\cos\omega_0\tau \:d\tau,
\qquad \Pi(t)=\int_0^t\kappa(\tau)\sin \omega_0\tau \:d\tau.
\end{equation}

In view of Eq. \eqref{scalar},  $\mathbf{X}^S$ and $\mathbf{P}^S$
do commute. As a consequence, $\mathbf{D}_S(t)$ can be regarded as
a quadratic form in the commutative operator variables
$\mathbf{X}^S$ and $\mathbf{P}^S$. This means that, using the
matrix notation introduced in Eq. \eqref{vnotation}, we can write
\begin{equation}
\label{vf}
\mathbf{D}_S(t)=(\vec{\mathbf{Z}}^{S})^t\:\mathsf{M}(t)\:\vec{\mathbf{Z}}^{S},
\qquad \text{with}\qquad \mathsf{M}(t)=
\begin{pmatrix}
\bar{\Delta}(t) & -\frac{\Pi(t)}{2}\\
-\frac{\Pi(t)}{2} & 0
\end{pmatrix}.
\end{equation}

It is not difficult to  check that
$\left[\mathbf{D}_S(t),\mathbf{D}_S(t_1)\right]=0$, whatever $t$
and $t_1$ are. Inserting Eq. \eqref{dissi} into Eq. \eqref{eqs1}
one can write the generalized non-Markovian Master Equation
describing the dissipative dynamics of our system in the following
final form \cite{rwa,leshouches}
\begin{equation}
\frac{d \hat{\rho}(t)}{dt} =  \left[\frac{1}{i \hbar}
\mathbf{\bar{H}}_{0}^S(t) -\mathbf{D}_S(t)
+\gamma(t)\left(\mathbf{N}+2\right)\right]\hat{\rho}(t) ,
\label{eqs}
\end{equation}
with
\begin{equation}
\label{rino} \hat{\bar{H}}_0(t)=
\frac{\hbar\omega_0}{2}\left[\hat{P}^2+\hat{X}^2-
\frac{r(t)}{\omega_0}\hat{X}^2
+\frac{\gamma(t)}{\omega_0}\left(\hat{X}\hat{P}+\hat{P}\hat{X}\right)\right].
\end{equation}
Let us note, first of all, that such a Master Equation is local in
time, even if non-Markovian. This feature is typical of all the
generalized Master Equations derived by using the
time-convolutionless projection operator technique
\cite{tcl,petruccionebook} or equivalent approaches such as the
superoperatorial one sketched in this section.

\begin{figure}
\includegraphics[width=5 cm,height=5 cm]{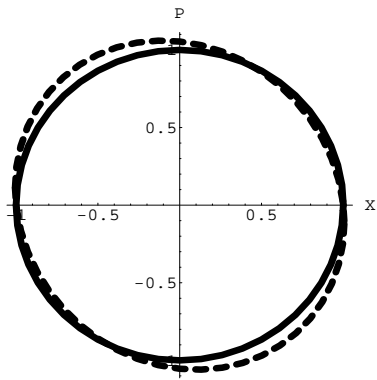}
\caption{\footnotesize The plain line represents the unperturbed
oscillator. The dashed line represents, instead, an oscillator
with an hamiltonian of the form given in Eq. \eqref{rino} (we have
chosen $\frac{\gamma(t)}{\omega_0}=\frac{r(t)}{\omega_0}=0.1$ )}
\label{ellipse}
\end{figure}

Let us have a closer look at the form of our Master Equation. Eq.
\eqref{rino} shows the appearance of two terms, produced by the
interaction with the environment, modifying the free hamiltonian
of the system. The first one,  proportional to $r(t)$, is a time
dependent renormalization of the frequency of the oscillator while
the second one, proportional to $\gamma(t)$, describes a coupling
between the the $\hat{X}$ and $\hat{P}$ operators. The
perturbation of the free dynamics due to these terms can be easily
visualized in phase space, as shown in Fig. (\ref{ellipse}).
Indeed, the effect of the terms proportional to $r(t)$ and
$\gamma(t)$ is equivalent to a compression and rotation in phase
space of the circle describing the free oscillator dynamics.
From a dynamical point of view one can show that the term
proportional to $\gamma(t)$ gives rise to both a further frequency
renormalization and  a dynamical dephasing  between position and
 momentum of the oscillator (see Fig. \ref{ellipse}). Note that these
 features are  analogue to the ones present in the dynamics of a classical
dissipative oscillator \cite{leshouches}.

\section{Operatorial Solution}

In quantum mechanics there exists a well established procedure to
determine the dynamical evolution of a given closed physical
system. It prescribes to write down the Schr\"{o}dinger equation
and calculate eigenstates and eigenvalues of the Hamiltonian of
the system. Once determined these quantities it is possible to
describe the dynamics evaluating the action of the corresponding
temporal evolution operator on the initial state. The dynamics of
open systems, however, is much more complicated since the
description of the state of the system in terms of a state vector
is not sufficient anymore. In order to describe completely the
physical system one needs indeed to introduce the density matrix
operator whose time evolution is governed by a master equation
definitely more difficult to handle than the Schr\"{o}dinger
equation.

In this section we present a new method to solve the generalized
Master equation, given by Eq.\eqref{eqs}, for the system of
interest in the paper. In some sense, the method we describe can
be seen as a generalization to open systems of the procedure for
solving the Schr\"{o}dinger equation of a closed system, since it
is based on the solution of an appropriate eigenvalue equation.
Indeed we will deal, in the formalism of superoperator, with
generalized eigenvalue equations involving superoperators and
operators, instead of operators and vectors respectively. As we
will see in this section, the existence of some useful algebraic
properties of the superoperators will help us in treating the
problem of the open system dynamics allowing, in particular, to
find the operatorial solution for the density matrix of the
system.

A remarkable virtue of  the procedure we are going to describe is
its independence on the expression of the time-dependent
coefficients appearing in the Master Equation \eqref{eqs}. This
implies that it is applicable to all the Master Equations
presenting the same structure.

Our method for solving the Master Equation  \eqref{eqs} consists
of two steps which can be summarized as follows:

%\begin{itemize}
%\item
1)  Singling out the temporal evolution superoperator
corresponding to Eq.\eqref{eqs};
%\item

2)  Understanding how this temporal evolution superoperator acts
on $\hat{\rho}(0)$.
%\end{itemize}

\subsection{Temporal Evolution Superoperator}

In the previous section we have introduced the Master Equation
related to the dissipative non-Markovian dynamics of our system:
\begin{equation}
\frac{d \hat{\rho}(t)}{dt} =  \left[\frac{1}{i \hbar}
\mathbf{\bar{H}}_{0}^S(t) -\mathbf{D}_S(t)
+\gamma(t)\left(\mathbf{N}+2\right)\right]\hat{\rho}(t).
\label{eqs2}
\end{equation}
A formal solution of Eq. \eqref{eqs2} can be written as
\begin{equation}
\hat{\rho}(t)=\mathbf{T}(t)\hat{\rho}(0) \label{rhot},
\end{equation}
where the temporal evolution superoperator $\mathbf{T}(t)$ is
given by
\begin{equation}
\label{tesup} \mathbf{T}(t)=exp_c\left(-\int_0^t\left[\frac{1}{i
\hbar} \mathbf{\bar{H}}_{0}^S(t_1) -\mathbf{D}_S(t_1)
+\gamma(t_1)\left(\mathbf{N}+2\right)\right]\:dt_1 \right).
\end{equation}
In \eqref{tesup} the pedix \lq\lq c\rq\rq stands for the Dyson
chronological order. In the following we shall prove that
$\mathbf{T}(t)$ can be factorized as follows:
\begin{equation}
\mathbf{T}(t)=\mathbf{T}_{Os}(t)\:\mathbf{T}_{\Gamma}(t)\:\mathbf{T}_D(t),
\label{fact}
\end{equation}
where
\begin{subequations}
\label{fact2}
\begin{eqnarray}
\mathbf{T}_{Os}(t)&=&exp_c\left[\frac{1}{i \hbar}\int_0^t
\mathbf{\bar{H}}_{0}^S(t_1)\:dt_1\right], \\
\mathbf{T}_{\Gamma}(t)&=&
exp\left[\frac{\Gamma(t)}{2}\left(\mathbf{N}+2\right)\right],\\
\mathbf{T}_D(t)&=&exp\left[-\int_0^t
e^{\Gamma(t_1)}\mathbf{\bar{D}}(t_1)\:dt_1\right],
\end{eqnarray}
\end{subequations}
with
\begin{eqnarray}
\Gamma(t)&=& 2\int_0^t \gamma(t_1)\:dt_1, \\
\mathbf{\bar{D}}(t)&=&\mathbf{T}_{Os}^{-1}(t)\:\mathbf{D}_S(t)\:\mathbf{T}_{Os}(t).
\label{ds}
\end{eqnarray}
To this aim we exploit the  Feynman's rule \cite{vank} stating
that, whatever the operator or superoperator $A(t)$ and $B(t)$
are,
\begin{equation}
exp_c\left[ \int \left( A(t)+B(t) \right)\:dt\right]=exp_c\left[
\int  A(t)\:dt\right]\:exp_c\left[ \int \bar{B}(t)\:dt\right],
\end{equation}
with
\begin{equation}
\bar{B}(t)=\left(exp_c\left[ \int
A(t)\:dt\right]\right)^{-1}\:B(t)\:\left(exp_c\left[ \int
A(t)\:dt\right]\right).
\end{equation}
Applying the  Feynman's rule to the time evolution superoperator
defined in Eq. \eqref{tesup} and taking in consideration that, as
shown in Appendix A, $ \left[\mathbf{N},
\mathbf{\bar{H}}_{0}^S(t)\right]=0$, one gets
\begin{equation}
\mathbf{T}(t)=\mathbf{T}_{Os}(t)\:\mathbf{T}'(t), \label{fact1}
\end{equation}
where
\begin{equation}
\mathbf{T}'(t)=exp_c\left[\int_0^t
\left[\gamma(t_1)\left(\mathbf{N}+2\right)-\mathbf{\bar{D}}(t_1)
\right]\:dt_1\right]. \label{f2}
\end{equation}
Applying again Feynman's rule in Eq.\eqref{f2} and using the
property (see Appendix A)
$\left[\mathbf{N},\mathbf{\bar{D}}(t)\right]=-2\mathbf{\bar{D}}(t)$,
leading to
\begin{equation}
\mathbf{T}^{-1}_{\Gamma}(t)\:\mathbf{\bar{D}}(t)\:\mathbf{T}_{\Gamma}(t)=e^{\Gamma(t)}\mathbf{\bar{D}}(t),
\end{equation}
we have
\begin{equation}
\label{factt2}
\mathbf{T}'(t)=\mathbf{T}_{\Gamma}(t)\:\mathbf{T}_D(t).
\end{equation}
Finally, inserting Eq.\eqref{factt2} into Eq. \eqref{fact1} one
obtains the final factorized form of the time evolution
superoperator given by Eq. \eqref{fact}.

The key advantage of such a factorized form is that it makes
easier estimating the action of the time evolution superoperator
on the initial  density matrix of the system. As we shall see in
the following, this allows not only to solve the Master Equation
\eqref{eqs}, but also to clarify the physical origin of each of
its terms.

It is worth noting that the Dyson chronological order is present
in only one of the three superoperators defined Eqs. \eqref{fact},
\eqref{fact2}. Of course, this circumstance leads to a further
simplification of the calculations. In fact, while for
$\mathbf{T}_{\Gamma}(t)$ the reason why one can drop the subscript
\lq\lq c\rq\rq is already clear from Eq. \eqref{fact2}, in the
case of $\mathbf{T}_D(t)$ some more comment is needed. As we will
demonstrate in the following, the Dyson chronological order is not
necessary in the expression of $\mathbf{T}_D(t)=exp\left[-\int_0^t
e^{\Gamma(t_1)}\mathbf{\bar{D}}(t_1)\:dt_1\right]$ because the
superoperator $\mathbf{\bar{D}}(t)$, defined in Eq.\eqref{ds}, is
a quadratic form in $\vec{\mathbf{Z}}^S$, that is
\begin{equation}
\label{dquadratic}
\mathbf{\bar{D}}(t)=(\vec{\mathbf{Z}}^{S})^t\:\mathsf{A}(t)\:\vec{\mathbf{Z}}^{S},
\end{equation}
with $\mathsf{A}(t)$ $2 \times 2$ matrix of time dependent scalar
quantities. Indeed it is straightforward to prove that, when Eq.
\eqref{dquadratic} holds, $\left[\mathbf{\bar{D}}(t),
\mathbf{\bar{D}}(t_1) \right]= 0$, for all $t, t_1$.

In order to prove Eq.\eqref{dquadratic}, let us note that, as
demonstrated in Appendix B, the following chain of equalities hold
\begin{equation}
\label{tran}
\mathbf{T}^{-1}_{Os}(t)\:\vec{\mathbf{Z}}^S\:\mathbf{T}_{Os}(t)=
\left(\hat{T}^{-1}_{Os}(t)\hat{\vec{Z}}\hat{T}_{Os}(t)\right)^S=
\left(\mathbf{T}^{-1}_{Os}(t)\:\hat{\vec{Z}}\right)^S.
\end{equation}
According to these equations, the transformation operated by
$\mathbf{T}_{Os}(t)$ on $\vec{\mathbf{Z}}^S$ is a linear
transformation correspondent to the one operated by
$\hat{T}_{Os}(t)$ on $\hat{\vec{Z}}$. One can show (see Appendix
B) that
\begin{equation}
\label{evolu}
\hat{T}_{Os}^{-1}(t)\:\hat{\vec{Z}}\:\hat{T}_{Os}(t)=\mathsf{R}(t)\:\hat{\vec{Z}},
\qquad \text{where}\qquad \mathsf{R}(t)=
\begin{pmatrix}
c(t) & s(t)\\
-s_r(t)& c_r(t)
\end{pmatrix}.
\end{equation}
The functions $c(t)$and $s(t)$ are the solutions of the Cauchy
problems
\begin{equation}
\label{equationcs}
\ddot{y}+\omega_0^2[1-\frac{r(t)}{\omega_0}-\frac{\gamma^2(t)}{\omega_0^2}-\frac{\dot{\gamma}(t)}{\omega_0^2}]y=0
\qquad\text{with initial conditions}\qquad
\begin{cases}
c(0)=1\qquad \dot{c}(0)=0\\
s(0)=0\qquad \dot{s}(0)=\omega_0
\end{cases}
\end{equation}
The functions $\dot{s}_r(t)$and $\dot{c}_r(t)$ are defined as
\begin{equation}
\label{csr}
c_r(t)=\frac{\dot{s}(t)}{\omega_0}-\frac{\gamma(t)}{\omega_0}s(t)
\qquad
-s_r(t)=\frac{\dot{c}(t)}{\omega_0}-\frac{\gamma(t)}{\omega_0}c(t).
\end{equation}
Inserting Eq. \eqref{vf} into Eq. \eqref{ds} and using Eqs.
\eqref{tran} one gets
\begin{equation}
\label{disegnato}
\mathbf{\bar{D}}(t)=(\vec{\mathbf{Z}}^{S})^t\:\mathsf{\bar{M}}(t)\:\vec{\mathbf{Z}}^{S},\qquad
\text{where we have put}\qquad
\mathsf{\bar{M}}(t)=\mathsf{R}^t(t)\mathsf{M}(t)\mathsf{R}(t).
\end{equation}

These considerations allow to recast $\mathbf{T}_D(t)$ in the form
\begin{equation}
\label{mathTd}
\mathbf{T}_D(t)=exp\left[-(\vec{\mathbf{Z}}^{S})^t\:\mathsf{W}(t)\:\vec{\mathbf{Z}}^{S}\right],
\end{equation}
where
\begin{equation}
\label{wuditi} \mathsf{W}(t)=\int_0^t
e^{\Gamma(t_1)}\mathsf{\bar{M}}(t_1)\:dt_1.
\end{equation}
Summarizing, in this subsection we have manipulated the expression
of the time evolution superoperator in order to put it in a form
more convenient for the calculations. In fact we have proved that
such a superoperator can be factorized in the form given by Eq.
\eqref{fact}. Moreover we have shown that  two of the
superoperators appearing in Eq. \eqref{fact} do not need the Dyson
chronological order. As we will see in the next section, this
circumstance allows to find an analytic expression for the density
matrix solution.

\subsection{Evolution of the density matrix}

In this subsection we calculate the action of the factorized
superoperator given in Eq. \eqref{fact} on the initial density
matrix.

Having in mind Eqs. \eqref{rhot} and \eqref{fact}, and writing the
initial density matrix as follows (see Eq. \eqref{ofe}):
\begin{equation}
 \hat{\rho}(0)=\frac{1}{2\pi}\int
 \chi_0(\vec{z'})\:e^{-\imath\:\vec{z'}^t\mathsf{J}\hat{\vec{Z}}}\:d^2\vec{z'},
 \label{rhozero}
\end{equation}
with
$\chi_0(\vec{z'})=\Tr{e^{\imath\:\vec{z'}^t\mathsf{J}\hat{\vec{Z}}}\hat{\rho}(0)}$,
we have
\begin{equation}
\rho(t)= \frac{1}{2 \pi} \int \mathbf{T}_{O_s}(t)
\mathbf{T}_{\Gamma}(t) \mathbf{T}_D(t)
\chi_0(\vec{z'})\:e^{-\imath\:\vec{z'}^t\mathsf{J}\hat{\vec{Z}}}\:d^2\vec{z'}.
\label{rhot2}
\end{equation}
Let us firstly introduce two superoperator eigenvalue equations we
will use in the following. Indicating with $F(\mathbf{A})$ a
generic superoperatorial well defined function of the
superoperator $\mathbf{A}$, from Eqs. \eqref{scalar}, \eqref{auto}
and \eqref{n} we obtain:
\begin{eqnarray}
F(\vec{\mathbf{Z}}^S)\:e^{-\imath\:\vec{z}^t\mathsf{J}\hat{\vec{Z}}}=
F(-\vec{z})\:e^{-\imath\:\vec{z}^t\mathsf{J}\hat{\vec{Z}}},
\label{property1} \\ F(\mathbf{N}) \left(
\vec{z}^t\mathsf{J}\hat{\vec{Z}} \right)^n =  F(n) \left(
\vec{z}^t\mathsf{J}\hat{\vec{Z}} \right)^n, \label{property2}
\end{eqnarray}
where now $F(-\vec{z})$ and $F(n)$ are simple scalar functions.

We begin noting that, using Eqs. \eqref{mathTd} and
\eqref{property1}, one can write the term $ \mathbf{T}_D(t)
\:e^{-\imath\:\vec{z'}^t\mathsf{J}\hat{\vec{Z}}}$ appearing in Eq.
\eqref{rhot2} as follows
\begin{equation}
\mathbf{T}_D(t) \:e^{-\imath\:\vec{z'}^t\mathsf{J}\hat{\vec{Z}}}=
e^{-(\vec{z'})^t\:\mathsf{W}(t)\:\vec{z'}}
\:e^{-\imath\:\vec{z'}^t\mathsf{J}\hat{\vec{Z}}}. \label{passo1}
\end{equation}
Moreover, using Eq. \eqref{property2} it is not difficult to prove
that
\begin{equation}
\mathbf{T}_{\Gamma}(t)\:e^{-\imath\:\vec{z}^t\mathsf{J}\hat{\vec{Z}}}=
e^{\Gamma(t)}\:\sum_n\frac{[e^{\frac{\Gamma(t)}{2}}(-\imath\:\vec{z}\,^t\mathsf{J}\hat{\vec{Z}})]^n}{n!}=
e^{\Gamma(t)}exp[e^{\frac{\Gamma(t)}{2}}(-\imath\:\vec{z}^t\mathsf{J}\hat{\vec{Z}})].
\label{passo2}
\end{equation}
Finally, remembering Eqs. \eqref{tran} and \eqref{evolu} we get
\begin{equation}
\mathbf{T}_{Os}(t)\:e^{-\imath\:\vec{z}^t\mathsf{J}\hat{\vec{Z}}}=
e^{-\imath\:\vec{z}^t\mathsf{J}\mathbf{T}_{Os}(t)\hat{\vec{Z}}}=
e^{-\imath\:\vec{z}^t\mathsf{J}\mathsf{R}^{-1}(t)\:\hat{\vec{Z}}}=
e^{-\imath\:(\mathsf{R}(t)\vec{z})^t\mathsf{J}\hat{\vec{Z}}}=
e^{-\imath\:\vec{z}(t)^t\mathsf{J}\hat{\vec{Z}}} \label{passo3}
\end{equation}
Inserting Eqs. \eqref{passo1}, \eqref{passo2} and \eqref{passo3}
into Eq. \eqref{rhot2} we can express the density matrix solution
as follows
\begin{equation}
\label{spro} \hat{\rho}(t)=\frac{1}{2\pi}\int
e^{-(\vec{z'})^t\:\mathsf{W}(t)\:\vec{z'}}
\chi_0(\vec{z'})\:e^{\Gamma(t)}exp[e^{\frac{\Gamma(t)}{2}}(-\imath\:\vec{z'}(t)^t\mathsf{J}\hat{\vec{Z}})]\:d^2\vec{z'}
\end{equation}

After some algebraic manipulation, reported in Appendix D, the
previous expression can be
 recast in the following final form

\begin{equation}
\label{sdef} \hat{\rho}(t)=\frac{1}{2\pi}\int
e^{-(\vec{z})^t\:\bar{\mathsf{W}}(t)\:\vec{z}}
\chi_0\left[e^{-\frac{\Gamma(t)}{2}}\mathsf{R}^{-1}(t)\:\vec{z}\right]\:
e^{-\imath\:\vec{z}^t\mathsf{J}\hat{\vec{Z}}}\:d^2\vec{z}=\frac{1}{2\pi}\int
\chi_t(\vec{z})\:
e^{-\imath\:\vec{z}^t\mathsf{J}\hat{\vec{Z}}}\:d^2\vec{z},
\end{equation}
where
\begin{equation}
\label{ancoraw}
 \bar{\mathsf{W}}(t)=e^{-\Gamma(t)}(\mathsf{R}^{-1}(t))^t
\:\mathsf{W}(t)\:\mathsf{R}^{-1}(t).
\end{equation}
Eq.\eqref{sdef} constitutes the main result of the paper. It gives
the operatorial density matrix solution of the problem of the
dissipative dynamics of a harmonic oscillator interacting with a
generic environment satisfying properties 2. and 3. of section I,
as for example a thermal reservoir at $T$ temperature. Our
approach to the dynamics of the system relies on the weak coupling
limit but does not invoke the Born--Markov and rotating wave (RW)
approximations. In more detail, we have solved a non-Markovian
generalized Master Equation for the harmonic oscillator coincident
with that deducible using either the time convolutionless
projection operator technique \cite{tcl} or the superoperatorial
approach of \cite{rwa}. At the best of our knowledge this is the
first analytical method allowing to solve such a generalized
Master Equation. In the next section we derive some approximated
forms of the solution given by Eq.\eqref{sdef} and we show its
usefulness in calculating the analytic expression of many
observables of interest and thus in gaining new insight in the
dynamics of one of the most extensively studied physical systems.

\section{Approximated forms of the solution and applications}

In this section we derive an approximated form of the operatorial
density matrix solution, given by Eq. \eqref{sdef},  valid when
\begin{equation}
\label{aiuuuto}
\hat{\bar{H}}_0(t)\approx\hat{H}_0,
\end{equation}
with $\hat{H}_0$ given by Eq. \eqref{rino}.  Looking at this
equation one sees immediately that Eq. \eqref{aiuuuto} amounts at
neglecting the time dependent frequency renormalization and
dephasing terms
\begin{equation}
\frac{r(t)}{\omega_0},\frac{\gamma^2(t)}{\omega_0^2}.
\end{equation}
It is worth noting that such an approximation is always justified
in the weak coupling regime $\alpha \ll 1$, provided that the
reservoir frequency cut--off remains finite, as one can appreciate
with the help of Eqs. \eqref{eq:secI16b}, \eqref{rtgt} and
\eqref{rino}. For the same reason it turns out that
$\frac{\dot{\gamma}(t)}{\omega_0}\ll1$. From
Eq.\eqref{equationcs}, it is not difficult to prove that this last
inequality allows to write the matrix $\mathsf{R}(t)$, defined in
Eq. \eqref{evolu}, as follows
\begin{equation}
\mathsf{R}(t)\approx
\begin{pmatrix}
\cos \omega_0t &\sin\omega_0t\\
-\sin\omega_0t& \cos \omega_0t
\end{pmatrix}.
\end{equation}
Inserting this expression in the definition of
$\bar{\mathsf{W}}(t)$, given by Eq. \eqref{ancoraw}, and
exploiting Eqs. \eqref{wuditi}, \eqref{disegnato} and \eqref{vf}
we get
\begin{equation}
\label{nrwa} \bar{\mathsf{W}}(t)\approx e^{-\Gamma(t)}\int_0^t
e^{\Gamma(t_1)}\left[\frac{\bar{\Delta}(t_1)}{2}+\frac{\bar{\Delta}(t_1)}{2}
\mathsf{C}_2(t-t_1)
-\frac{\Pi(t_1)}{2}\mathsf{S}_2(t-t_1)\right]dt_1,
\end{equation}
with
\begin{equation}
\mathsf{C}_2(t)=\begin{pmatrix}
\cos 2\omega_0t & -\sin 2\omega_0 t\\
-\sin 2\omega_0t & -\cos 2\omega_0t
\end{pmatrix},\qquad
\mathsf{S}_2(t)=
\begin{pmatrix}
\sin 2\omega_0t& \cos 2\omega_0t\\
\cos 2\omega_0t& -\sin 2\omega_0t
\end{pmatrix}.
\end{equation}

The form of the matrices defined in the previous equations
suggests a further approximation very common in Quantum Optics:
the Rotating Wave Approximation (RWA). It basically consists in
neglecting rapidly oscillating terms as, in our case, those
oscillating at frequency $2\omega_0$. This amounts at averaging to
zero all the elements of the matrices $\mathsf{C}_2(t)$ and
$\mathsf{S}_2(t)$ so that Eq. \eqref{nrwa} reduces to
\begin{equation}
\label{ancora} \bar{\mathsf{W}}(t)\approx e^{-\Gamma(t)}\int_0^t
e^{\Gamma(t_1)}\frac{\bar{\Delta}(t_1)}{2}dt_1=\frac{\Delta_{\Gamma}(t)}{2}.
\end{equation}
Substituting Eq. \eqref{ancora} into Eq. \eqref{sdef} we obtain
the following expression for the density matrix solution in the
RWA
\begin{equation}
\label{sappr} \hat{\rho}(t)\approx\hat{\rho}^{\rm RWA
}(t)=\frac{1}{2\pi}\int
e^{-\frac{\Delta_{\Gamma}(t)}{2}\abs{\vec{z}}^2}
\chi_0\left[e^{-\frac{\Gamma(t)}{2}}\mathsf{R}^{-1}(t)\:\vec{z}\right]\:
e^{-\imath\:\vec{z}^t\mathsf{J}\hat{\vec{Z}}}\:d^2\vec{z}.
\end{equation}
It is possible to demonstrate that $\hat{\rho}^{\rm RWA }(t)$, as
given by Eq. \eqref{sappr}, satisfies the following Master
Equation
\begin{equation}
\frac{d}{dt} \hat{\rho}^{\rm RWA }(t) =  \left[\frac{1}{i \hbar}
\mathbf{H}_{0}^S
-\frac{\bar{\Delta}(t)}{2}\abs{\vec{\mathbf{Z}}^S}^2
+\gamma(t)\left(\mathbf{N}+2\right) \right]\hat{\rho}^{\rm RWA
}(t) , \label{arwa}
\end{equation}
which in turn has been derived in \cite{rwa}.
%which, without the vectorial representation and in the interaction
%picture, can be rewritten as

%\begin{equation}
%\frac{d}{dt} \hat{\rho}^{\rm RWA }(t) = - \left[
%\frac{\bar{\Delta}(t)}{2}\left[(\mathbf{X}^S)^2+(\mathbf{P}^S)^2\right]
%+\frac{\imath\gamma(t)}{2}\left(\mathbf{X}^{S}\mathbf{P}^{\Sigma}-\mathbf{P}^{S}\mathbf{X}^{\Sigma}\right)
%\right]\hat{\rho}^{\rm RWA }(t) , \label{arwa}
%\end{equation}

Once obtained the density matrix we are able, at least in
principle, to evaluate the mean value $\med{A}$ of each and every
operator $A$ of interest for the system. One of the advantages of
having a solution of the density matrix in terms of the
characteristic function is the possibility of exploiting the
following relations \cite{gardiner}:
\begin{equation}
\label{xp}
\med{\hat{X}^n}=(-\imath)^n\left(\frac{\partial^n}{\partial p^n
}\chi(x,p)\right)_{x,p=0},\qquad
\med{\hat{P}^n}=(\imath)^n\left(\frac{\partial^n}{\partial x^n
}\chi(x,p)\right)_{x,p=0}.
\end{equation}
By using this relations it is not difficult to calculate the time
evolution of the mean energy of the oscillatory system:
\begin{equation}
\label{energy}
\med{\hat{H}_0}_t=e^{-\Gamma(t)}\med{\hat{H}_0}_{t=0}+\hbar
\omega_0
\begin{cases}
\Delta_{\Gamma}(t)&\text{with RWA,}\\
\Tr{\bar{\mathsf{W}}(t)}&\text{without RWA but neglecting
renormalization and dephasing terms.}
\end{cases}
\end{equation}
We emphasize that these solutions depend on the initial state of
the oscillator only through the term $\med{\hat{H}_0}_{t=0}$, the
second term of the sum being independent of the initial state.

Moreover, looking at Eq. \eqref{energy} one sees that the
exponential factor accounting for energy dissipation does not
depend on the initial state of the system but only on the
characteristic parameters of the reservoir.\\
These features, characterizing  Eq. \eqref{energy}, are directly
related to the factorized form of the the QCF $\chi_t(\vec{z})$.
In fact, from both Eq. \eqref{sdef} and
 Eq. \eqref{sappr} it appears evident that such a function is the
product of a gaussian factor and the function
$\chi_0\left[e^{-\frac{\Gamma(t)}{2}}\mathsf{R}^{-1}(t)\:\vec{z}\right]$.
All the information on the initial state of the system is
incorporated in this last function from which the first term in
the righthand side of Eq. \eqref{energy} comes from. On the
contrary, the second  term derives exclusively from the gaussian
factor of the QCF, which depends only on the environment functions
and  not on the initial state of the system.

Note from Eqs. \eqref{sdef} and \eqref{sappr} that, whatever  the
initial state of the system is,
\begin{equation}
\chi_t(\vec{z})\xrightarrow{t\rightarrow
\infty}e^{-(\vec{z})^t\:\bar{\mathsf{W}}(t)\:\vec{z}}\simeq{\text{(in
the RWA) }}e^{-\frac{\Delta_{\Gamma}(t)}{2}\abs{\vec{z}}^2}
\end{equation}
This behavior is easily understandable, when the environment is
 a thermal reservoir, in the light of the
thermalization process.

Let us now have a closer look at Eq. \eqref{energy}. If we
evaluate $\Tr{\bar{\mathsf{W}}(t)}$ with the help of Eq.
\eqref{nrwa}, we find that
\begin{equation}
\Tr{\bar{\mathsf{W}}(t)}=\Delta_{\Gamma}(t).
\end{equation}
This means that the time evolution of the oscillator energy is not
affected by the contribution of the rapidly oscillating terms
neglected in the RWA. This feature comes directly from the
particular structure of the free Hamiltonian operator.

To better understand this point we consider the gaussian factor
appearing in the QCF. The superoperator corresponding to such a
factor has the form
\begin{equation}
\mathbf{T}_G(t)=exp[-(\vec{\mathbf{Z}}^S)^t\:\bar{\mathsf{W}}(t)\:\vec{\mathbf{Z}}^S].
\end{equation}
With some algebraic manipulation one can recast the superoperator
appearing in the exponent as follows
\begin{equation}
(\vec{\mathbf{Z}}^S)^t\:\bar{\mathsf{W}}(t)\:\vec{\mathbf{Z}}^S=\frac{\Delta_{\Gamma}(t)}{2}\left[(\mathbf{X}^S)^2+(\mathbf{P}^S)^2\right]
+\frac{\Lambda(t)}{2}\left[(\mathbf{X}^S)^2-(\mathbf{P}^S)^2\right]+\Theta(t)\mathbf{X}^S\mathbf{P}^S,
\end{equation}
where the time dependent coefficients appearing in the previous
equation are
\begin{equation}
\Delta_{\Gamma}(t)=\Tr{\bar{\mathsf{W}}(t)},\qquad
\Lambda(t)=\Tr{\hat{\sigma}_z\bar{\mathsf{W}}(t)},\qquad
\Theta(t)=\Tr{\hat{\sigma}_x\bar{\mathsf{W}}(t)},
\end{equation}
with $\hat{\sigma}_z$ and $\hat{\sigma}_x$  Pauli spin matrices.

Exploiting the properties of the trace and putting
$\hat{\rho}'(t)=\mathbf{T}_G(t)^{-1}\hat{\rho}(t)$, one can easily
show that the following chain of equalities holds
\begin{equation}
\label{tpe}
\med{\hat{A}}_t=\Tr{\hat{\rho}(t)\hat{A}}=\Tr{\mathbf{T}_G(t)\hat{\rho}'(t)\hat{A}}=
\Tr{\hat{\rho}'(t)\mathbf{T}_G(t)\hat{A}}=\Tr{\hat{\rho}'(t)\hat{A}^G(t)},
\end{equation}
where we have defined
\begin{equation}
\hat{A}^G(t)\equiv\mathbf{T}_G(t)\hat{A}.
\end{equation}
Having these equations in mind it is not difficult to convince
oneself that calculating $\med{\hat{A}}_t$ using $\chi_t(x,p)$ is
equivalent at calculating $\med{\hat{A}^G}_t$ using
$\chi_0\left[e^{-\frac{\Gamma(t)}{2}}\mathsf{R}^{-1}(t)\:\vec{z}\right]$.
In the case
$\hat{A}=\hat{H}_0=\frac{1}{2}\left(\hat{P}^2+\hat{X}^2\right)$ we
have that
\begin{equation}
\left[(\mathbf{X}^S)^2+(\mathbf{P}^S)^2\right]\hat{H}_0=-2\hbar
\omega_0,
\qquad\left[(\mathbf{X}^S)^2-(\mathbf{P}^S)^2\right]\hat{H}_0=0,
\qquad \mathbf{X}^S\mathbf{P}^S\hat{H}_0=0,
\end{equation}
and thus
\begin{equation}
\label{hog} \hat{H}_0^G(t)=\hat{H}_0+\hbar
\omega_0\Delta_{\Gamma}(t).
\end{equation}
Note that this equation is not affected by the RWA as well as
$\chi_0\left[e^{-\frac{\Gamma(t)}{2}}\mathsf{R}^{-1}(t)\:\vec{z}\right]$.
This explains why we obtain for the mean value of the oscillator
energy the same result with or without the RWA approximation.

The previous procedure suggests a sufficient condition to single
out operators which \lq \lq do not suffer the RWA
approximation\rq\rq, indeed we have that
\begin{equation}
\label{condition}
\left[(\mathbf{X}^S)^2-(\mathbf{P}^S)^2\right]\hat{A}=0, \qquad
\mathbf{X}^S\mathbf{P}^S\hat{A}=0\qquad\Rightarrow
\med{\hat{A}}_t=\med{\hat{A}}_t^{\rm RWA}.
\end{equation}
With the expression operators which \lq \lq do not suffer the RWA
approximation\rq\rq we indicate operators having the property that
the time evolution of their mean value is not influenced by the
RWA. In other words the counter rotating terms do not contribute
to the dynamics of this class of observables. Examples of
operators belonging to such a class are $\hat{X}$, $\hat{P}$,
$\hat{X}^2+\hat{P}^2$ and all linear combinations of such
operators, as one can easily verify. On the contrary
 $\hat{X}^2$ and $\hat{P}^2$
do not satisfy condition \eqref{condition}. For such operators,
indeed, exploiting Eq. \eqref{tpe},  one gets
\begin{subequations}
\begin{equation}
\med{\hat{X}^2}_t-\med{\hat{X}^2}_t^{\rm RWA}=-\Lambda(t),\qquad
\med{\hat{P}^2}_t-\med{\hat{P}^2}_t^{\rm RWA}=\Lambda(t).
\end{equation}
Similarly, for the mean value of the \lq\lq correlation\rq\rq
operator $(\hat{X}\hat{P}+\hat{P}\hat{X})$ one has
\begin{equation}
\med{\hat{X}\hat{P}+\hat{P}\hat{X}}_t-\med{\hat{X}\hat{P}+\hat{P}\hat{X}}_t^{\rm
RWA}=2 \Theta(t).
\end{equation}
\end{subequations}

\section{Conclusions}

In this paper we have developed a method to solve the weak
coupling generalized Master Equation for QBM. In particular we
have considered the case in which the system interacting with the
environment is a quantum harmonic oscillator. The Master Equation
we have solved, given by Eq.\eqref{eqs}, although non--Markovian
is local in time. Such a Master Equation has been derived in
\cite{rwa} by using a superoperatorial technique and in
\cite{petruccionebook,leshouches} by means of the time
convolutionless method and is the weak coupling approximated form
of the exact Master Equation for QBM derived by Paz and Zurek in
\cite{MEexact}. The main result of the paper is the derivation of
the analytic solution of Eq.\eqref{eqs} for the density matrix of
the reduced system. To this aim we have used an approach based on
the algebra of superoperators. Our method is independent of the
specific form of the environment and does not rely on any
approximation apart from the weak coupling one. We have also
studied  simpler forms of the density matrix solution obtained
neglecting frequency renormalization terms and/or performing the
RWA. We have demonstrated the existence of a class of
superoperators whose mean value is not affected by the presence of
the counter rotating terms at any time $t$ and we have given a
sufficient condition to verify if a given operator belongs to such
a class. This circumstance simplifies substantially the
calculations since, for operators belonging to such a class, one
can use the approximated density matrix solution, given by
Eq.\eqref{sappr}, in order to calculate their mean value. The
analytic solution we derive and discuss in the paper is given in
terms of the QCF (see Eqs.\eqref{sdef}) by means of which one can
calculate the expectation value of many observable of physical
interest in a very direct way, as suggested by Eqs.\eqref{xp}. For
example, thanks to the simplicity of the analytic solution we have
derived, we succeed in calculating the dissipative time evolution
of the mean energy of the system. Finally it is worth noting that
from the QCF it is easy to derive the Wigner function
characterizing the state of the dissipative system.

Concluding, we believe that, the non--Markovian analytic approach
we have derived in this paper for QBM can be generalized to other
fundamental dissipative systems, such as for example the
Jaynes--Cummings model with losses. As for the QBM, we think that
the analytic solution of the density matrix may be used for
studying new aspects of such a basic model, both for fundamental
and for applicative research, under conditions in which, up to
now, only numerical approaches were possible.

\section{Acknowledgements}
One of the authors (S.M.) acknowledges financial support from
Finanziamento Progetto Giovani Ricercatori anno 1999, Comitato 02.

\appendix

\section{Properties of dissipation superoperator}

In this Appendix we  demonstrate some useful properties of the
superoperator $\mathbf{N}$. Such properties, allowing to factorize
the time-evolution superoperator given in Eq \eqref{tesup}, are
the following
\begin{equation}
\left[\mathbf{N}, \mathbf{\bar{H}}_{0}^S(t)\right]=0, \qquad
\left[\mathbf{N},\mathbf{\bar{D}}(t)\right]=-2\mathbf{\bar{D}}(t).
\end{equation}

The first one is the most surprising one: it can be considered as
a \lq\lq dynamical invariance\rq\rq of the dissipation process. As
we shall see in the following,  the \lq\lq dynamical
invariance\rq\rq originates from the particular structure of the
Hamiltonian term $\hat{\bar{H}}_{0}(t)$, defined by Eq.
\eqref{rino}.

Independently on the form of the time dependent functions present
in such a term, it can be written, in general,  as
\begin{equation}
\hat{\bar{H}}_{0}(t)=\hat{\vec{Z}}^t \mathsf{S}(t)\hat{\vec{Z}}
\qquad\text{with} \qquad \mathsf{S}^t(t)=\mathsf{S}(t)
\end{equation}
where $\mathsf{S}(t)$ is a $2\times 2$ real symmetric matrix.
Remembering that
\begin{equation}
\mathbf{N}=-\frac{1}{2}\imath\:(\vec{\mathbf{Z}}^{\Sigma})^t\mathsf{J}\vec{\mathbf{X}^{S}},
\end{equation}

one gets
\begin{subequations}
\begin{eqnarray}
\label{inv} \left[\mathbf{N},\mathbf{\bar{H}}_{0}^S(t)\right]&=&
-\frac{1}{2}\imath\:
\left[(\vec{\mathbf{Z}}^{\Sigma})^t\mathsf{J}\vec{\mathbf{Z}}^{S},
\left(\hat{\vec{Z}}^t \mathsf{S}(t)\hat{\vec{Z}}\right)^S
\right]\nonumber\\
&=&\frac{1}{2}\imath\:\left(
(\vec{\mathbf{Z}}^{\Sigma})^t\mathsf{J}\left[\hat{\vec{Z}}^t
\mathsf{S}(t)\hat{\vec{Z}},\hat{\vec{Z}}\right]^S + \left(\left[
\hat{\vec{Z}}^t \mathsf{S}(t)\hat{\vec{Z}},\hat{\vec{Z}}
\right]^{\Sigma}\right)^t\mathsf{J}\vec{\mathbf{Z}}^{S} \right)
\end{eqnarray}
It is possible to show that \cite{goldstein}
\begin{equation}
\label{der} \left[\hat{\vec{Z}}^t
\mathsf{S}(t)\hat{\vec{Z}},\hat{\vec{Z}}\right]=\imath
\mathsf{J}\mathsf{S}(t)\hat{\vec{Z}}.
\end{equation}
Substituting this equation into Eq. \eqref{inv} and using the
properties of the matrices $\mathsf{S}(t)$ and $\mathsf{J}$ (cfr
Eq. \eqref{matrixj}) one obtains

\begin{eqnarray}
\left[\mathbf{N},\mathbf{\bar{H}}_{0}^S(t)\right]&=&
-\frac{1}{2}\left(
(\vec{\mathbf{Z}}^{\Sigma})^t\mathsf{J}\mathsf{J}\mathsf{S}(t)\vec{\mathbf{Z}}^S
+
(\vec{\mathbf{Z}}^{\Sigma})^t\mathsf{S}^t(t)\mathsf{J}^t\mathsf{J}\vec{\mathbf{Z}}^{S}
\right)\nonumber\\
&=& -\frac{1}{2}\left(
-(\vec{\mathbf{Z}}^{\Sigma})^t\mathsf{S}(t)\vec{\mathbf{Z}}^S +
(\vec{\mathbf{Z}}^{\Sigma})^t\mathsf{S}(t)\vec{\mathbf{Z}}^{S}
\right)=0.
\end{eqnarray}
\end{subequations}

The generality of the matrix $\mathsf{S}(t)$ ensures the validity
of this relation for a generic quadratic hamiltonian and thus, in
particular, for $\hat{\bar{H}}_{0}(t)$.

As far as the second commutation relation is conserned,
remembering that
$\mathbf{\bar{D}}(t)=(\vec{\mathbf{Z}}^{S})^t\:\mathsf{\bar{M}}(t)\:\vec{\mathbf{Z}}^{S}$
one has
\begin{subequations}
\begin{eqnarray}
\label{autod} \left[\mathbf{N},\mathbf{\bar{D}}(t)\right]&=&
\left[\mathbf{N},
(\vec{\mathbf{Z}}^{S})^t\:\mathsf{\bar{M}}(t)\:\vec{\mathbf{Z}}^{S}\right]\nonumber\\
&=& \left[\mathbf{N},
\vec{\mathbf{Z}}^{S}\right]^t\:\mathsf{\bar{M}}(t)\:\vec{\mathbf{Z}}^{S}+
(\vec{\mathbf{Z}}^{S})^t\:\mathsf{\bar{M}}(t)\:\left[\mathbf{N},
\vec{\mathbf{Z}}^{S}\right]
\end{eqnarray}
Exploiting the superoperatorial commutation rules given in Eq.
\eqref{comrel} yields
\begin{equation}
\left[\mathbf{N},\vec{\mathbf{Z}}^{S}\right]=
-\frac{\imath}{2}\left[\left(\mathbf{P}^{\Sigma}\mathbf{X}^{S}-\mathbf{X}^{\Sigma}\mathbf{P}^{S}\right),
\vec{\mathbf{Z}}^{S}\right]=-\vec{\mathbf{Z}}^{S}.
\end{equation}
Finally, substituting into Eq. \eqref{autod} one gets
\begin{equation}
\left[\mathbf{N},\mathbf{\bar{D}}(t)\right]=-(\vec{\mathbf{Z}}^{S})^t\:\mathsf{\bar{M}}(t)\:\vec{\mathbf{Z}}^{S}-
(\vec{\mathbf{Z}}^{S})^t\:\mathsf{\bar{M}}(t)\:\vec{\mathbf{Z}}^{S}=-2\mathbf{\bar{D}}(t)
\end{equation}
\end{subequations}

\section{Time evolution}
In this Appendix we shall demonstrate that the general time
evolution of a \lq\lq S\rq\rq- or \lq\lq $\Sigma$\rq\rq-type
superoperator is equivalent to the \lq\lq S\rq\rq- or \lq\lq
$\Sigma$\rq\rq-type superoperator of the time evolution of the
corresponding operator. In formulas this amounts at demonstrating
Eq. \eqref{tran}. First of all let us define $T_c(t)$ as solution
of the equation
\begin{equation}
\label{tc} \frac{d}{dt}T_c(t)=L(t)T_c(t)
\end{equation}
where in general $[L(t),L(t_1)]\not=0$. $L(t)$, as well as
$T_c(t)$, can be either an operator or a superoperator.

The previous equation can be solved in iterative way and its
solution is
\begin{equation}
\label{scexp}
T_c(t)=\sum_{n=0}\int_0^t\cdots\int_0^{t_{n-1}}L(t_1)\cdots
L(t_n)\:dt_1\cdots dt_n\equiv exp_c\left[\int_0^t
L(t_1)\:dt_1\right]
\end{equation}
It can be shown that $T^{-1}_c(t)$ satisfies the equation
\cite{vank}
\begin{equation}
\label{ta} \frac{d}{dt}T^{-1}_c(t)=-T^{-1}_c(t)L(t)
\end{equation}
which can be again solved in iterative way giving the following
form of the solution
\begin{equation}
\label{saexp}
T^{-1}_c(t)=\sum_{n=0}(-1)^n\int_0^t\cdots\int_0^{t_{n-1}}L(t_n)\cdots
L(t_1)\:dt_1\cdots dt_n\equiv exp_a\left[-\int_0^t
L(t_1)\:dt_1\right]
\end{equation}

Let us define
\begin{equation}
A_c(t)=T_c(t)\:A\:T^{-1}_c(t)\equiv \mathcal{T}_c(t)\:A
\end{equation}
From the previous equation we have
\begin{equation}
\frac{d}{dt}{\cal{T}}_c(t)\:A=[L(t),A_c(t)]=L^S(t)A_c(t)=L^S(t)\mathcal{T}_c(t)\:A
\qquad \Rightarrow
\frac{d}{dt}\mathcal{T}_c(t)=L^S(t)\mathcal{T}_c(t)
\end{equation}
This means that $\mathcal{T}_c(t)$ satisfies an equation similar
to the one given by Eq. \eqref{tc}. The form of $\mathcal{T}_c(t)$
can be obtained from Eq \eqref{scexp} replacing $T_c(t)\rightarrow
\mathcal{T}_c(t)$ and $L(t)\rightarrow L^S(t)$.

Now, let us suppose that $A$ is \lq\lq S\rq\rq- or \lq\lq
$\Sigma$\rq\rq-type superoperator and $L$ is an \lq\lq
S\rq\rq-type superoperator. From the properties given in
\eqref{comrel} we have that
\begin{eqnarray}
\label{B7}
\mathbf{A}^{S(\Sigma)}_c(t)&=&\mathbf{T}_c(t)\:\mathbf{A}^{S(\Sigma)}\:\mathbf{T}^{-1}_c(t)=
\mathcal{T}_c(t)\mathbf{A}^{S(\Sigma)}\nonumber\\
&=&\sum_{n=0}\int_0^t\cdots\int_0^{t_{n-1}}[\mathbf{L}^S(t_1)\cdots
[\mathbf{L}^S(t_n),\mathbf{A}^{S(\Sigma)}]\cdots]\:dt_1\cdots dt_n\nonumber\\
&=&\sum_{n=0}\int_0^t\cdots\int_0^{t_{n-1}}[\hat{L}(t_1)\cdots
[\hat{L}(t_n),\hat{A}]\cdots]^{S(\Sigma)}\:dt_1\cdots dt_n\nonumber\\
&=&\left(\sum_{n=0}\int_0^t\cdots\int_0^{t_{n-1}}\mathbf{L}^S(t_1)\cdots
\mathbf{L}^S(t_n)\hat{A}\:dt_1\cdots dt_n \right)^{S(\Sigma)}\nonumber\\
&=&\left(exp_c\left[\int_0^t\mathbf{L}^S(t_1)\:dt_1\right]\hat{A}\right)^{S(\Sigma)}=
\left(\mathbf{T}_c(t)\:\hat{A}\right)^{S(\Sigma)}\nonumber\\
&=&\left(\hat{T}_c(t)\:\hat{A}\:\hat{T}^{-1}_c(t)\right)^{S(\Sigma)}=\left(\hat{A}_c(t)\right)^{S(\Sigma)}
\end{eqnarray}
where we have put $\hat{T}_c(t)\equiv
exp_c\left[\int_0^t\hat{L}^S(t_1)\:dt_1\right]$.

Now, let us consider
\begin{equation}
\mathbf{A}^{S(\Sigma)}_a(t)=\mathbf{T}^{-1}_c(t)\:\mathbf{A}^{S(\Sigma)}\:\mathbf{T}_c(t)\equiv
\mathcal{T}^{-1}_c(t)\mathbf{A}^{S(\Sigma)}.
\end{equation}
The last definition comes directly from the definition of
$\mathcal{T}_c(t)$ and can be verified applying $\mathcal{T}_c(t)$
on $\mathbf{A}^S_a(t)$ and, viceversa, applying the previous
definition of $\mathcal{T}^{-1}_c(t)$ on $\mathbf{A}^S_c(t)$.\\
Exploiting Eq.\eqref{saexp} and following the same lines of the
derivation of Eq. \eqref{B7}, one can show that
\begin{equation}
\mathbf{A}^{S(\Sigma)}_a(t)
=\left(\hat{T}^{-1}_c(t)\:\hat{A}\:\hat{T}_c(t)\right)^{S(\Sigma)}=\left(\hat{A}_a(t)\right)^{S(\Sigma)}.
\end{equation}

Let us now derive  Eq. \eqref{evolu}. Using Eq. \eqref{der},
\eqref{tc}, \eqref{ta} and remembering the definition of
$\hat{\vec{Z}}(t)$:
\begin{equation}
\hat{\vec{Z}}(t)=\hat{T}_{Os}^{-1}(t)\:\hat{\vec{Z}}\:\hat{T}_{Os}(t).
\end{equation}
one hase
\begin{equation}
\label{der1}
\frac{d}{dt}\hat{\vec{Z}}(t)=-\frac{1}{\imath\hbar}\hat{T}_{Os}^{-1}(t)\:[\hat{\bar{H}}_0(t),\hat{\vec{Z}}]\:\hat{T}_{Os}(t)
=-\mathsf{J}\mathsf{\tilde{S}}(t)\hat{\vec{Z}}(t)\qquad\mathsf{\tilde{S}}(t)=\frac{1}{\hbar}\mathsf{S}(t).
\end{equation}
Deriving the previous equation once more we get
\begin{equation}
\frac{d^2}{dt^2}\hat{\vec{Z}}(t)=
\left[-\mathsf{J}\mathsf{\dot{\tilde{S}}}(t)+(\mathsf{J}\mathsf{\tilde{S}}(t))^2\right]\hat{\vec{Z}}(t)=
\left[-\mathsf{J}\mathsf{\dot{\tilde{S}}}(t)-det[\mathsf{\tilde{S}}(t)]\right]\hat{\vec{Z}}(t).
\end{equation}

In our case
\begin{equation}
\mathsf{\tilde{S}}(t)=\frac{1}{2}
\begin{pmatrix}
\omega_0-r(t)&\gamma(t)\\
\gamma(t)&\omega_0
\end{pmatrix},
\end{equation}
so that the differential equation for the operator $\hat{X}(t)$
has the following form
\begin{equation}
\frac{d^2}{dt^2}\hat{X}(t)=
-\omega_0^2\left(1-\frac{r(t)}{\omega_0}-\frac{\gamma^2(t)}{\omega^2_0}-\frac{\dot{\gamma}(t)}{\omega^2_0}\right)\hat{X}(t).
\end{equation}
A solution of the previous equation can be written as
\begin{equation}
\label{x(t)} \hat{X}(t)=\hat{X}c(t)+\hat{P}s(t),
\end{equation}
with $c(t)$ and $s(t)$ solutions of Eq. \eqref{equationcs} with
the same initial conditions. Note that, from \eqref{der1}, it
follows that
\begin{equation}
\frac{d}{dt}\hat{X}(t)=\omega_0\hat{P}(t)+\gamma(t)\hat{X}(t).
\end{equation}
With th help of Eq. \eqref{x(t)} we have for $\hat{P}(t)$
\begin{eqnarray}
\label{p(t)}
\hat{P}(t)&=&\frac{1}{\omega_0}\left(\frac{d}{dt}\hat{X}(t)-\gamma(t)\hat{X}(t)\right)\nonumber\\
&=&\hat{P}\frac{\dot{s}(t)-\gamma(t)s(t)}{\omega_0}+
\hat{X}\frac{\dot{c}(t)-\gamma(t)s(t)}{\omega_0\nonumber}\\
&=&\hat{P}c_r(t)-\hat{X}s_r(t),
\end{eqnarray}
where we have defined, as in Eq. \eqref{csr},
\begin{equation}
c_r(t)=\frac{\dot{c}(t)-\gamma(t)s(t)}{\omega_0}\qquad
-s_r(t)=\frac{\dot{c}(t)-\gamma(t)s(t)}{\omega_0}.
\end{equation}
Grouping Eqs. \eqref{x(t)}, \eqref{p(t)} and using the matrix
representation we obtain
\begin{equation}
\begin{cases}
\hat{X}(t)=\hat{X}c(t)+\hat{P}s(t)\\
\hat{P}(t)=-\hat{X}s_r(t)+\hat{P}c_r(t)
\end{cases}
\Rightarrow\hat{\vec{Z}}(t)=\mathsf{R}(t), \qquad \mathsf{R}(t)=
\begin{pmatrix}
c(t)&s(t)\\
-s_r(t)&c_r(t)
\end{pmatrix}.
\end{equation}

\section{Final form of the density matrix}

In this appendix we sketch the main steps of the derivation of the
final form of the density matrix solution, given by Eq.
\eqref{sdef}, from Eq. \eqref{spro}. Let us remind the generic
expression for the density matrix in terms of the QCF (See Eq.
\eqref{ofe})
\begin{equation}
\label{ofet} \hat{\rho}(t)=\frac{1}{2\pi}\int \chi_t(\vec{z})\:
e^{-\imath\:\vec{z}^t\mathsf{J}\hat{\vec{Z}}}\:d^2\vec{z},
\end{equation}
where
$\chi_t(\vec{z})=\Tr{e^{\imath\:\vec{z}^t\mathsf{J}\hat{\vec{Z}}}\hat{\rho}(t)}$.
From Eq.\eqref{spro} we have
\begin{equation}
\label{sproa} \hat{\rho}(t)=\frac{1}{2\pi}\int
e^{-(\vec{z'})^t\:\mathsf{W}(t)\:\vec{z'}}
\chi_0(\vec{z'})\:e^{\Gamma(t)}exp[e^{\frac{\Gamma(t)}{2}}(-\imath\:\vec{z'}(t)^t\mathsf{J}\hat{\vec{Z}})]\:d^2\vec{z'}
\end{equation}
Comparing these last two equations one gets

\begin{equation}
\label{ex}
\chi_t(\vec{z})=\Tr{e^{\imath\:\vec{z}^t\mathsf{J}\hat{\vec{Z}}}\hat{\rho}(t)}=\frac{1}{2\pi}\int
e^{-(\vec{z'})^t\:\mathsf{W}(t)\:\vec{z'}}
\chi_0(\vec{z'})\:e^{\Gamma(t)}
\Tr{e^{\imath\:\vec{z}^t\mathsf{J}\hat{\vec{Z}}}exp[e^{\frac{\Gamma(t)}{2}}(-\imath\:\vec{z'}(t)^t\mathsf{J}\hat{\vec{Z}})]}\:d^2\vec{z'}
\end{equation}

From Eq.\eqref{ort} and using some properties of the $delta$
function we obtain
\begin{equation}
\frac{1}{2\pi}\Tr{e^{\imath\:\vec{z'}^t\mathsf{J}\hat{\vec{Z}}}
exp[e^{\frac{\Gamma(t)}{2}}(-\imath\:\vec{z}(t)^t\mathsf{J}\hat{\vec{Z}})]}=
\delta[\vec{z}-e^{\frac{\Gamma(t)}{2}}\vec{z'}(t)]=
e^{-\Gamma(t)}\delta[\vec{z'}-e^{-\frac{\Gamma(t)}{2}}\mathsf{R}^{-1}(t)\:\vec{z}].
\end{equation}

Introducing such this equation into Eq. \eqref{ex} yields

\begin{eqnarray}
\chi_t(\vec{z})&=&\int e^{-(\vec{z'})^t\:\mathsf{W}(t)\:\vec{z'}}
\chi_0(\vec{z'})
\delta[\vec{z'}-e^{-\frac{\Gamma(t)}{2}}\mathsf{R}^{-1}(t)\:\vec{z}]
\:d^2\vec{z'}\nonumber\\
&=&exp[-e^{-\Gamma(t)}(\vec{z})^t\:(\mathsf{R}^{-1}(t))^t\mathsf{W}(t)\mathsf{R}^{-1}(t)\:\vec{z}]
\chi_0[e^{-\frac{\Gamma(t)}{2}}\mathsf{R}^{-1}(t)\:\vec{z}]
\end{eqnarray}

Finally, substituting this equation into Eq.\eqref{ofet} gives

\begin{equation}
\label{sdefa} \hat{\rho}(t)=\frac{1}{2\pi}\int
e^{-(\vec{z})^t\:\bar{\mathsf{W}}(t)\:\vec{z}}
\chi_0\left[e^{-\frac{\Gamma(t)}{2}}\mathsf{R}^{-1}(t)\:\vec{z}\right]\:
e^{-\imath\:\vec{z}^t\mathsf{J}\hat{\vec{Z}}}\:d^2\vec{z},
\end{equation}
where
\begin{equation}
%\label{ancoraw}
 \bar{\mathsf{W}}(t)=e^{-\Gamma(t)}(\mathsf{R}^{-1}(t))^t
\:\mathsf{W}(t)\:\mathsf{R}^{-1}(t).
\end{equation}
which is the form of $\hat{\rho}(t)$ given in Eq.\eqref{sdef}.

\end{document}